# Fano Resonant Optical coatings platform for Full Gamut and High Purity Structural Colors

Mohamed ElKabbash[1,2, †, *,] Nathaniel Hoffman[3, †], Andrew R Lininger[3, †], Sohail A. Jalil[1], Theodore Letsou[3], Michael Hinczewski[3, *], Giuseppe Strangi[3, *], and Chunlei Guo[1, *]

1. The Institute of Optics, University of Rochester, Rochester, NY 14627, USA.
2. Current address: Research Laboratory of Electronics, MIT, Cambridge, MA, 02139, USA
3. Department of Physics, Case Western Reserve University, 10600 Euclid Avenue, Cleveland, Ohio 44106, USA.

* Corresponding emails: melkabba@mit.edu (M.E.), mxh605@case.edu (M.H.), gxs284@case.edu (G.S.), guo@optics.rochester.edu (C.G.).

†These authors contributed equally.

**Abstract**

Structural coloring is a photostable and environmentally friendly coloring approach that harnesses optical interference and nanophotonic resonances to obtain colors with a range of applications including steganography, décor, data storage, and anticounterfeiting measures. We show that optical coatings exhibiting the photonic Fano Resonance is an ideal platform for structural coloring- it provides full color access, high color purity, high brightness, controlled iridescence, and scalable manufacturability. We show that an additional oxide film deposited on Fano resonant optical coatings (FROCs) increases the color purity (up to 97%) and color gamut coverage range (> 99% coverage of the sRGB and Adobe color spaces). For coloring applications that do not require high spatial resolution, FROCs provide a significant advantage over existing structural coloring schemes.

**Introduction:**

In nature, colors are mostly produced either through pigments or structures. While the former comes from molecular absorption, the latter, structural coloring (SC), is the result of optical interference from a structured surface. Structural colors offer several advantages over pigments- they are photostable, immune to chemical degradation, and environmentally friendly. In addition, a wide range of colors can be produced[1] and dynamically reconfigured [2]using the same material. Furthermore, structured surfaces can have multiple functionalities, e.g., creating hydrophobic or antibacterial colored metals[3,4].

Several structural coloring schemes have been previously introduced including multilayer films[5], thin film nanocavities[6], plasmonic nanostructures [1], dielectric nanostructures[7], and photonic crystals[8] with applications in decoration[9], colorimetric sensing[10], data storage[11], anticounterfeiting[12], display technologies[13], colorful photovoltaic cells[14], among others[15]. An ideal structural coloring platform should span a wide color gamut, producing colors with high and controllable purity- how monochromatic or pure the color is- and brightness - the relative intensity of the reflected color- and allowing control over the colors' angle dependence. For many applications, it should also be scalable and inexpensive to fabricate. However, no existing scheme can satisfy all the above qualities simultaneously[1,7,9,14].

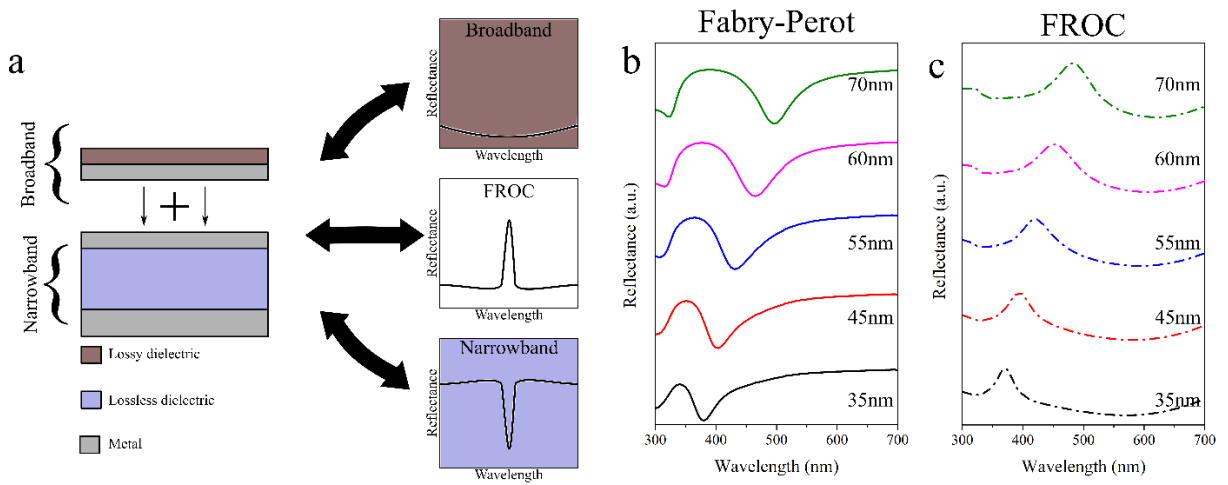

**Figure 1| Spectral properties of Fano Resonant Optical Coatings:** (a) A FROC consists of two coupled light absorbers; a broadband absorber and a narrowband (Fabry-Perot) absorber. A FROC exhibits a reflection peak at the Fabry-Perot cavity resonance. (b) The measured reflection from a Fabry-Perot cavity with different dielectric thickness. (c) The measured reflection from the same Fabry-Perot cavities shown in (b) after depositing a 15 nm Ge film to create a FROC. The incidence angle in (b) and (c) is 15°.

Recently, we proposed a new type of optical coatings that exhibits the photonic Fano Resonance effect [16]. Fano Resonant Optical Coatings (FROCs) enjoy unique optical properties that cannot be reproduced with existing optical coatings such as metallic films, anti-reflective coatings, transmission filters, light absorbers, and dielectric mirrors. **Figure 1a** describes the composition of FROCs. FROCs are produced by coupling a broadband nanocavity (representing the continuum) with a narrowband Fabry-Perot nanocavity (representing a discrete state). The resonant interference between the nanocavities produces the well-known asymmetric Fano resonance line-shape. In this work, we develop a class of reflective FROCs by using a reflective and opaque material as substrates, which lead to a highly reflective resonant peak that corresponds to the narrowband nanocavity's resonance. We investigate the color properties of FROCs numerically and experimentally. We show that the reflective FROCs are an ideal platform for structural coloring, producing colors spanning a wide color gamut with high brightness and high purity. The dependence of the color on the incident angle can be controlled through the cavity material, making FROCs suitable for a variety of applications that demand angle independence, e.g., decoration, or angle dependence, e.g., anti-counterfeit measures. Structural coloring with FROCs can find new applications where strong and broadband optical absorption and high purity colors are required, for example, colorful solar thermal generation panels, colorful photovoltaic panels[8], colorful thermophotovoltaic panels, and colorful solar thermoelectric generators. These renewable energy sources often have the same color, black or dark blue, which makes them aesthetically unappealing.

**Results and Discussion:**

Throughout this work, we compare the coloring performance of FROCs to Fabry-Perot (FP) thin film nanocavities since they are the closest platform in terms of structure and physics to FROCs [16]. While Fabry-Perot nanocavities produce colors through selective absorption, mainly reflecting all colors except for the specific cavity resonance wavelength. On the other hand, FROCs produce colors through selective reflection (**Fig. 1a**). **Figure 1b** shows the measured *p*-polarized reflection spectrum of Fabry-Perot nanocavities consisting of Ag (20 nm)-$TiO_2$-Ag (100 nm) by varying the thickness of the $TiO_2$ film from 35 nm to 70 nm. **Figure 1c** shows the measured reflection spectrum of the same Fabry-Perot nanocavities after adding a 15 nm Ge layer to convert them into FROCs. The reflection lines produced by the FROCs can span the visible spectrum and target a relatively narrow range of wavelengths for each individual FROC by changing the cavity's thickness.

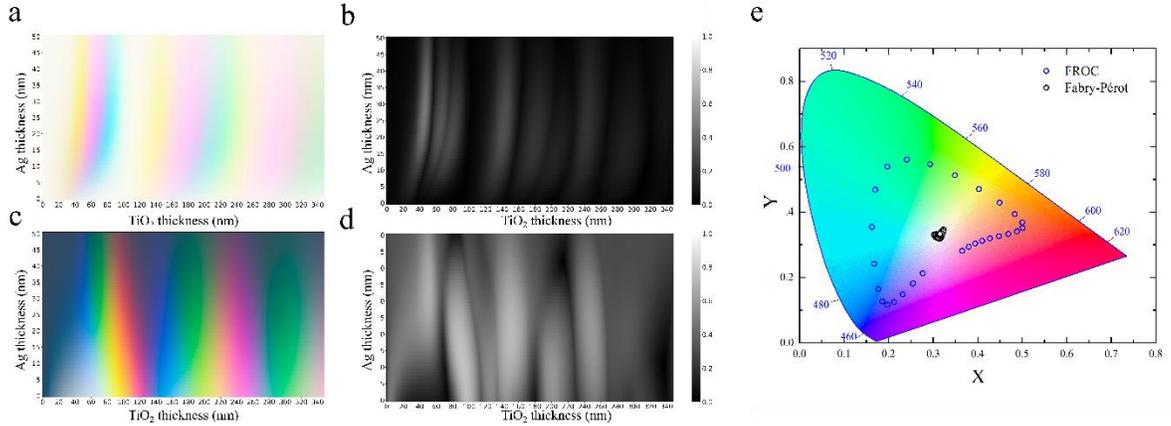

**Figure 2| Colorimetric properties of FROCs:** Swatch array and the corresponding color purity for Fabry-Perot (FP) nanocavities (a) and (b) and FROCs (c) and (d). (e) CIE 1931 color space showing the colors corresponding to the calculated reflection spectrum of FP nano-cavities (black circles) and FROC (blue circles) with varying cavity thicknesses.

To examine the structural coloring properties of FROCs vs. Fabry-Perot (FP) cavities, we calculate the colors produced from FP cavities vs. FROCs by varying the top metal film thickness and the cavity thickness. **Figure 2a** shows a swatch array for FP cavities consisting of a metal-dielectric-metal stack [top to bottom: Ag (Y nm)- $TiO_2$ (X nm)- Ag (100 nm)] as a function of the top Ag film thickness and the $TiO_2$ dielectric cavity thickness. A swatch is the perceived color for a person viewing the sample. The produced colors are Cyan-Magenta-Yellow (CMY) colors since FP cavities are selective absorbers and CMY colors are subtractive colors[17]. Note the restricted color palette from FP nanocavities. **Figure 2b** shows the corresponding purity of the FP cavities (see *Methods* for more details). The purity of FP cavities is limited since all the colors are reflected except within the absorbed wavelength range, making near-monochromatic reflection impossible. **Figure 2c** shows a swatch array of FROCs that are identical to the FP cavities with an added absorbing thin film [Ge (15 nm)] to generate the desired Fano resonance. The selective reflection from FROCs enables access to a wide range of hues from blue to red, including green. The colors enjoy significantly high purity as shown in **Figure 2d** [18]. The white point corresponds to the spectrum of the illuminant, i.e., white light. FROCs access the huge color gamut since they provide selective reflection at different wavelengths by simply changing the dielectric thickness.

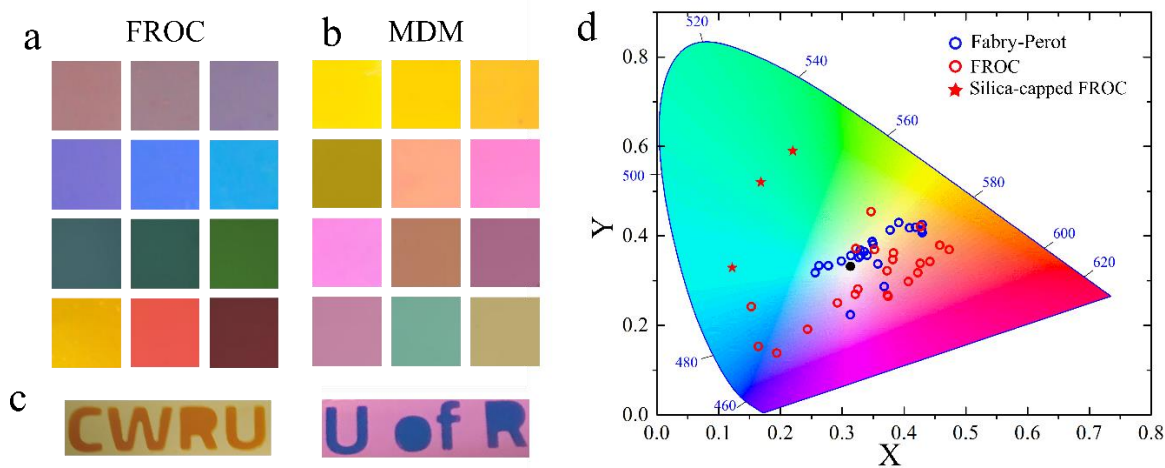

**Figure 3| Structural coloring with FROCs:** (**a**) and (**b**) Show photographs of fabricated FP nano-cavities and FROCs. The color purity of FROCs is evident in (**c**) where the letters of U of R and CWRU are printed on an MDM cavity by depositing 15 nm Ge layer. (**d**) The CIE 1931 color space showing the colors corresponding to the measured reflection spectrum of FP nano-cavities (blue circles) and FROCs (red circles). FROCs demonstrate higher purity as they are further away from the white point (black dot). Silica capped FROCs (red stars) show higher purity in the green region of the color space as we discuss later in the manuscript.

Images of the reflected color for FROC and MDM structures with a $TiO_2$ thickness ranging from 35 nm to 150 nm are shown in **Fig. 3a and Fig. 3b,** respectively. FROCs are capable of reflecting blue, green, and red colors by simply increasing the dielectric thickness. **Figure 3c** shows a photograph of two FP cavities with CWRU and U of R letters "printed" on them by depositing a 15nm Ge layer and converting these regions to a FROC. By controlling the spatial distribution of the deposited layer, this printing method could be adapted for optical archival data storage and encrypting messages. Experimental reflection lines from different FP cavities (blue dots) and FROCs (red dots) are presented in the CIE 1931 color space (see *Methods*) that links distributions of electromagnetic wavelengths to visually perceived colors (**Fig. 3d**). The experimental results agree with the numerical calculations of colors produced by FROCs vs. FP cavities (see Supplementary Information **Fig. S1**, **Fig. S2**, and **Fig. S3**). High purity green colors were difficult to obtain using conventional FROCs and were obtained using silica capped FROCs (red stars) as we will discuss below.

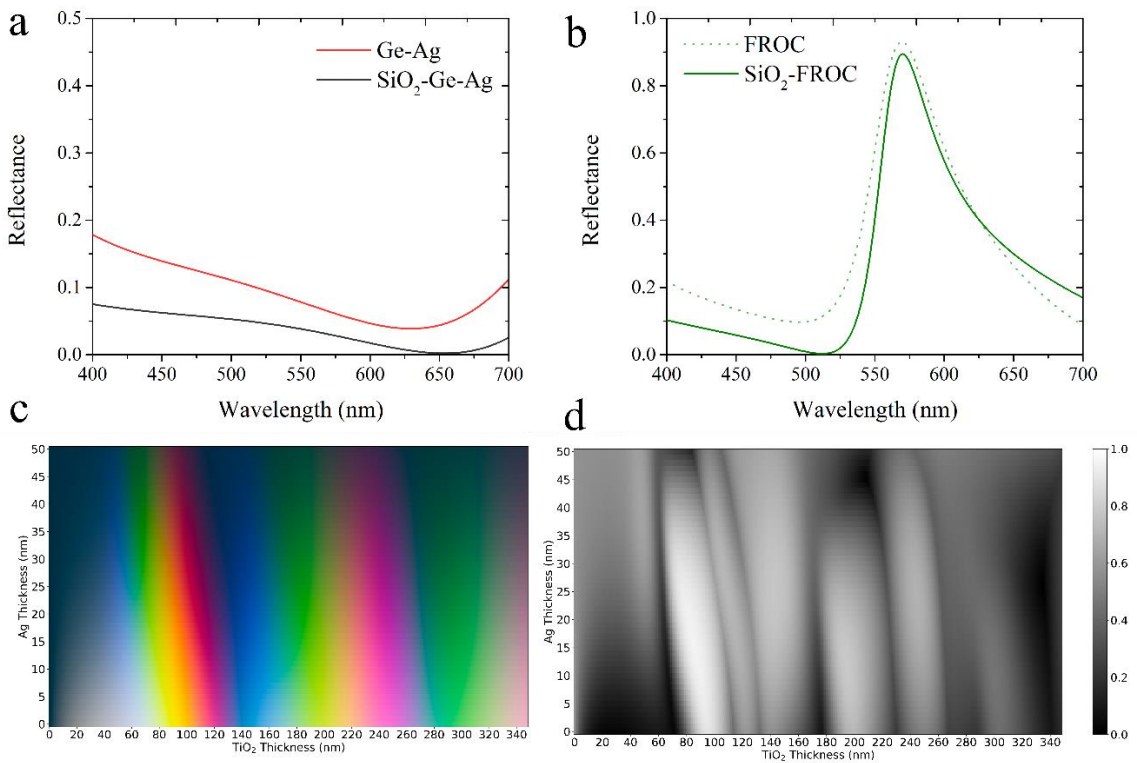

**Figure 4| Silica capped FROCs: (a)** Measured reflectance of a Ge-Ag broadband absorber (red line) vs. a silica capped broadband absorber (black line). Adding a silica film reduces the overall reflectance from the broadband absorber. **(b)** Measured reflectance of a FROC with a Fano resonance peak within the green wavelength range with and without a silica cap. The suppressed reflectance at shorter wavelengths is evident. **(c)** and **(d)** show a swatch array and color purity for silica capped FROCs ($SiO_2$ – 50 nm) for different thickness of the nanocavity and top Ag film.

High purity green and red colors are difficult to obtain. This agrees with the purity results and the corresponding colors of FROCs shown **Fig. 2c** and **Fig. 2d**. This is because obtaining red and green colors requires having high reflectance at longer wavelengths. However, the measured reflection from the Ge-Ag broadband absorber is > 0.15 at short wavelengths < 500 nm (**Fig. 4a**). Consequently, the color purity drops since other colors are reflected. To suppress the stronger reflection at shorter wavelengths, a 50nm silicon dioxide capping layer was added as shown in **Fig. 4a**. **Figure 4b** shows the measured reflection for a FROC that produces a green color with and without the $SiO_2$ capping layer. The $SiO_2$ capped FROC exhibits significantly reduced reflection at short wavelengths and relatively small change near the resonance peak. This combination leads to an overall increase in the colorimetric purity. **Fig. 4c** and **Fig. 4d** show the swatch array and corresponding color purity of the $SiO_2$ capped FROC design, respectively. By comparing **Fig. 4d** and **Fig. 2d**, a wider range of colors and greater purities can be obtained as compared to the original FROC structure. A purity level of > 97% can be reached in the yellow region of the spectra (Supplementary information, **Fig. S4 and Fig. S5**) which is significantly higher than other ultrahigh purity structural color platforms[7].

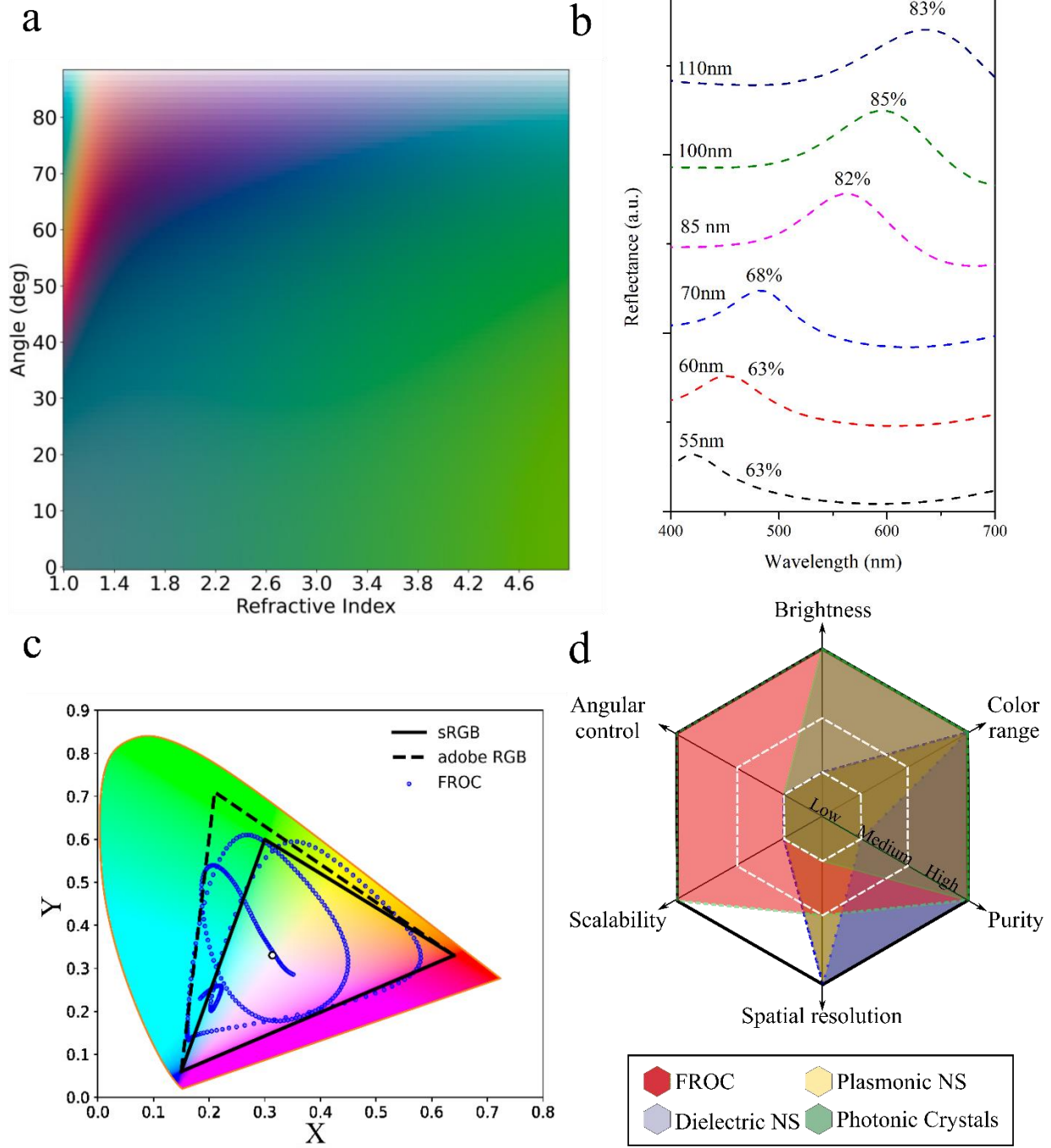

**Figure 5| Angle dependence, brightness, and accessible color gamut of FROCs: (a)** Swatch array of a silica capped FROC for a constant optical length by varying the refractive index and incidence angle. The angle dependence can be controlled by the refractive index of the dielectric cavity. **(b)** Measured reflectance from different FROCs showing a peak reflectance ranging from 0.63-0.85. The high reflectance corresponds to high brightness, a desirable property in structural colors. **(c)** CIE 1931 chromaticity diagram of Silica capped FROCs for different dielectric cavity thicknesses at normal (0°) incidence. The sRGB and Adobe RGB subspaces are shown for comparison. The FROC obtains 43% coverage of the total CIE 1931 diagram. The white point is shown as an open circle. **(d)** Comparison between the coloring performance of FROC vs. other structural coloring platforms.

In addition, FROC's iridescence can be controlled by simply changing the refractive index of the dielectric cavity. **Figure 5a** shows a swatch array (angle of incidence vs. cavity refractive index) for a range of FROCs with the same optical length $l = n\,t = 700\ nm$, where $n$ is the dielectric index and $t$ is its thickness, as a function of incidence angle. For high refractive indices, FROCs color over nearly the entire range of incident angles. Note that colors' angle independence may not be a desired property in some applications, e.g., anti-counterfeit measures.

Another important metric is brightness, i.e., the reflected intensity compared to the incident light at the peak resonance wavelength. The calculated reflectance from FROCs is > 0.9 (Supplementary information, **Fig. S2**) and the measured reflectance ranges from 0.63- 0.85 as shown in **Figure 5b**. The lower reflectance in the measured films can be improved by depositing higher quality silver films and precisely controlling thickness [19]. Finally, we assess the color range covered by silica capped FROCs. We obtain the CIE 1931 chromaticity coordinates of a stack of $SiO_2$ (50 nm) -Ge (15 nm)- Ag (48 nm)- $TiO_2$ (x nm)- Ag (100 nm) and vary the thickness of the $TiO_2$ layer. A total coverage of 42.9% is achieved at 0º viewing angle (**Figure 5c**). Adding a silica capping layer improves the covered color gamut of the bare FROC which is only 28% at 0 viewing angle. At 55º viewing angle, the covered percentage of the CIE color space is 60% coverage (**Supplementary information, Fig. S6**). We also calculate the coverage of the sRGB color space (the standard color space for the web), and the Adobe RGB color space (the color space that encompasses most of the colors achievable on CMYK color printers). The theoretical silica capped FROCs exhibit high coverage of both color spaces, with 89.3% of the sRGB color space and 85.5% of the Adobe color space, at normal incidence. The coverage percentage increases to > 99% coverage of both color spaces at 55º viewing angle (Supplementary information, **Fig. S7)**. Access to a wide color gamut with a similar reflection profile has been realized recently with multipolar metasurfaces, which - in contrast with FROCs- require intense nanolithography[7].

**Conclusions and Outlook:**

**Figure 5d** compares the structural coloring performance metrics of FROCs, dielectric nanostructures, plasmonic nanostructures, and photonic crystals. Indeed, FROCs outperform existing structural coloring methods by simultaneously offering high purity, access to a wide range of colors, high brightness, angular control, and cheap and scalable fabrication[1]. Because thin film structural coloring in general has lower spatial resolution compared to nanostructures based structural coloring, the latter remains advantageous for high density coloring. Durable FROCs can be made with ceramic materials [20]. We believe that FROCs are particularly suitable for colored solar thermal panels as they are efficient in absorbing the solar spectrum [16] while reflecting an on demand, narrowband color. In addition, by using amorphous Si instead of Ge and utilizing the metal films in the FROC to act as electrodes, it is possible to realize colorful photovoltaic cells[14] using the well-established thin film deposition technologies.

**Methods:**

**Sample fabrication:**

The FROC films were deposited on a glass substrate (Micro slides, Corning) using electron-beam evaporation for Ge (3 Å s$^{-1}$) and TiO$_2$ (1 Å s$^{-1}$) pellets and thermal evaporation Ag (20 Å s$^{-1}$), with the deposition rates specified for each material. The silica capped FROC films were deposited on a glass substrate (2948, Corning) using electron-beam evaporation for Ge (0.5 Å s$^{-1}$), TiO$_2$ (1 Å s$^{-1}$), and SiO$_2$ (0.8 Å s$^{-1}$), and DC magnetron sputtering for Ag (2 Å s$^{-1}$). All deposition materials were purchased from Kurt J. Lesker. Deposited layer thicknesses were measured with spectroscopic ellipsometry (J. A. Woollam).

**Numerical calculation of the reflection and absorption spectra:**

Numerical reflectance and absorbance spectra were generated using a transfer matrix-based simulation model written in Mathematica and Python. Spectral optical constants for the multiple materials were obtained variously from the Brendel-Bormann model (Ag), fits to the experimental materials (SiO$_2$, TiO$_2$), and an amorphous experimental model for Ge. Transmittance was zero for all structures. Absorbance was calculated as the complementary to reflectance, or $A = 1 - R$.

**Reflection measurements:** Experimental angular reflectance measurements were performed using a variable-angle high-resolution spectroscopic ellipsometer (V-VASE, J. A. Woollam). Sample transmittance was zero for all angles and wavelengths.

**Color analysis:**

The reflectance spectra to CIE 1931-xyz colorspace conversation was performed in Python utilizing interpolations of the standard observer distributions.[21,22] The XYZ tristimulus values are given as:

$$X = \frac{\int_\lambda S(\lambda)\,\alpha(\lambda)\,R(\lambda)\,d\lambda}{\int_\lambda S(\lambda)\,\beta(\lambda)\,d\lambda}, \quad Y = \frac{\int_\lambda S(\lambda)\,\beta(\lambda)\,R(\lambda)\,d\lambda}{\int_\lambda S(\lambda)\,\beta(\lambda)\,d\lambda}, \text{ and } Z = \frac{\int_\lambda S(\lambda)\,\gamma(\lambda)\,R(\lambda)\,d\lambda}{\int_\lambda S(\lambda)\,\beta(\lambda)\,d\lambda},$$

with $S$ as the illuminant spectrum, $R$ as the spectral reflectance, $k$ as a constant factor, and $\alpha, \beta, \gamma$ as the standard observer functions. The integration is over the visible spectrum. The CIE 1931-xyz values are then given as:

$$x = Y/N, \quad y = Y/N, \text{ and } z = 1 - x - y,$$

for $N = X + Y + Z$. Note that at constant luminance, the chromaticity is defined by $x$ and $y$.

Color swatch arrays were generated by transforming the calculated chromaticity values into their sRGB equivalents using a matrix transform calculated from reference primaries with the D65 reference white and sRGB companding (IEC 61966-2-1 standard). Colors were generated with matplotlib in Python.

Excitation purity is calculated as:

$$p = |s - w| / |d - w|$$

where $s$, $w$, and $d$ are the CIE 1931 (x, y) coordinates for the measured spectra point, white point, and dominant wavelength point, respectively.

Total CIE x-y space coverage is calculated as the area of the smallest convex hull encompassing all of the desired CIE (x, y) points. Area is presented relative to the area of the full visible light color gamut. Sufficient resolution was obtained in the numerical simulations to approach a smooth hull.

**Authors Contribution:**
M.E. developed the approach and designed the project. C.G., G. S., and M. H. supervised the project. N. H. and A. L. performed color analysis. A. L. and S. A. J. fabricated the samples. A. L., M. E., T. L. performed reflection measurements. M. E. wrote the manuscript with inputs from N. H. and A. L. All authors discussed the results.

**Acknowledgments**:
C. G acknowledges the support of the Army Research Office, The Bill and Melinda Gates foundation, and the National Science Foundation. G. S. and M. H. acknowledge the support of the National Science Foundation. This work made use of the High Performance Computing Resource in the Core Facility for Advanced Research Computing at Case Western Reserve University.

**Ethics declarations:**
Competing interests
A patent application has been filed on the Fano resonance optical coating scheme in this work.

**Data availability:**
Data are available upon request from the corresponding authors.

# Supplementary Information

# Fano Resonant Optical coatings platform for Full Gamut and High Purity Structural Colors

Mohamed ElKabbash[1,2, †, *,] Nathaniel Hoffman[3, †], Andrew R Lininger[3, †], Sohail A. Jalil[1], Theodore Letsou[3], Michael Hinczewski[3, *], Giuseppe Strangi[3, *], and Chunlei Guo[1, *]

1. The Institute of Optics, University of Rochester, Rochester, NY 14627, USA.
2. Current address: Research Laboratory of Electronics, MIT, Cambridge, MA, 02139, USA
3. Department of Physics, Case Western Reserve University, 10600 Euclid Avenue, Cleveland, Ohio 44106, USA.

* Corresponding emails: melkabba@mit.edu (M.E.), mxh605@case.edu (M.H.), gxs284@case.edu (G.S.), guo@optics.rochester.edu (C.G.).

†These authors contributed equally.

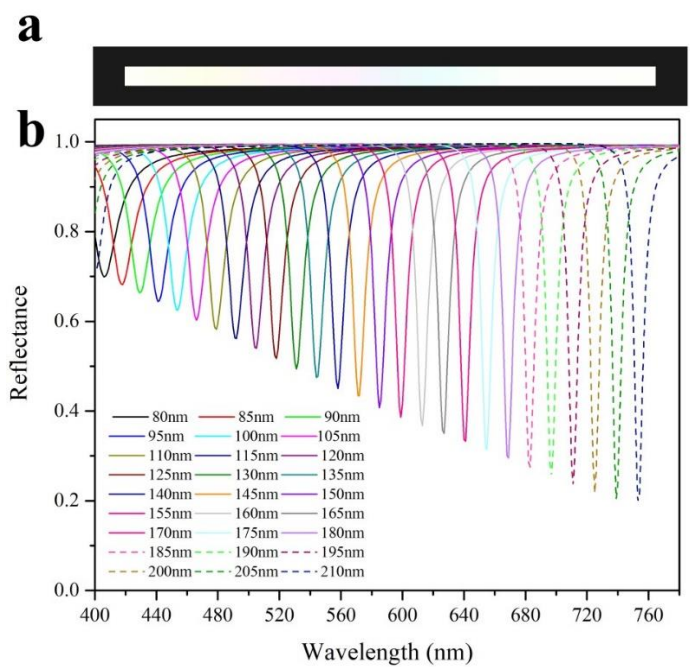

**Figure S1| Colors from Fabry-Perot MDM cavities: (a)** The calculated colors corresponding to MDM cavities by varying the dielectric ($SiO_2$) thickness from 80 nm to 210 nm. The narrow absorption lines shown in **(b)** lead to low purity colors.

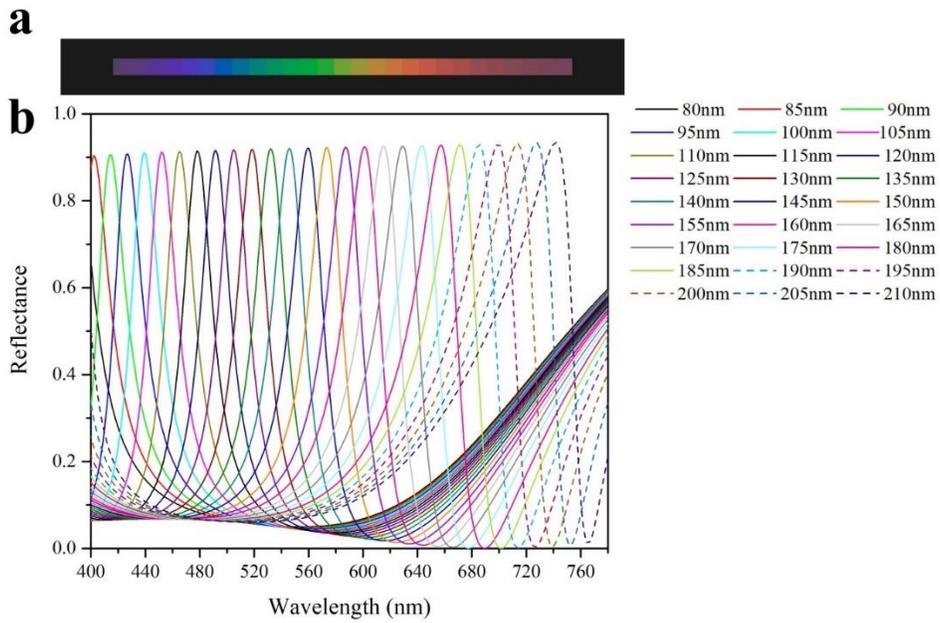

**Figure S2| Colors from FROCs: (a)** The calculated colors corresponding to FROCs cavity by varying the dielectric (SiO$_2$) thickness from 80 nm to 210 nm. The narrow reflection lines shown in **(b)** lead to high purity colors.

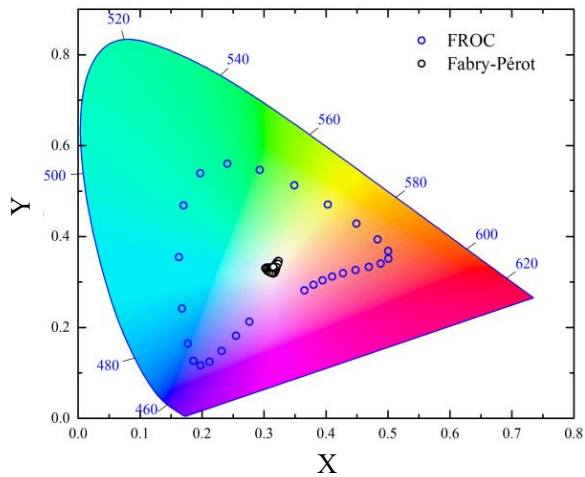

**Figure S3|** CIE 1931 color space showing the colors corresponding to the calculated reflection spectrum of FP nano-cavities (black circles) and FROC (blue circles) with varying cavity thicknesses.

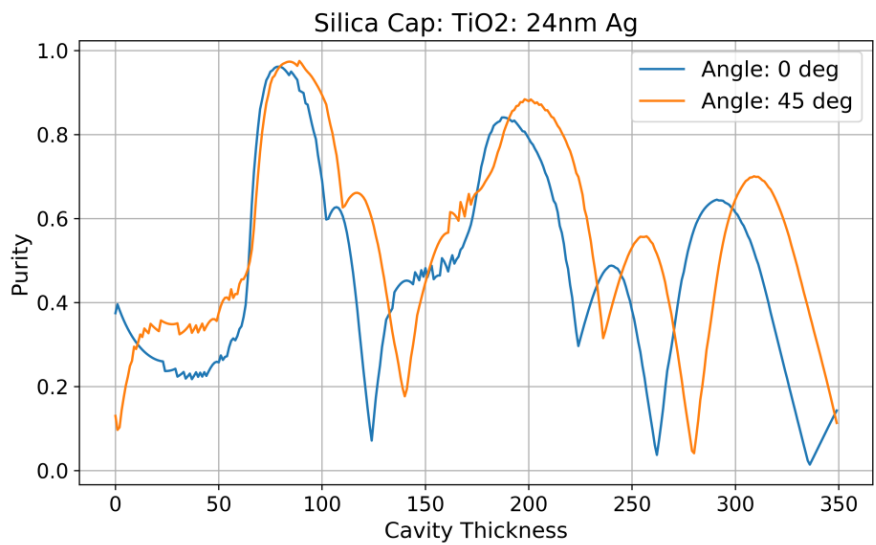

**Figure S4|** The color purity of Silica capped FROCs consisting of $SiO_2$ (50 nm) -Ge (15 nm)- Ag (24 nm)- $TiO_2$ (x nm)- Ag (100 nm). The $TiO_2$ thickness is varied. Purity levels > 97% can be achieved over a wide angular range.

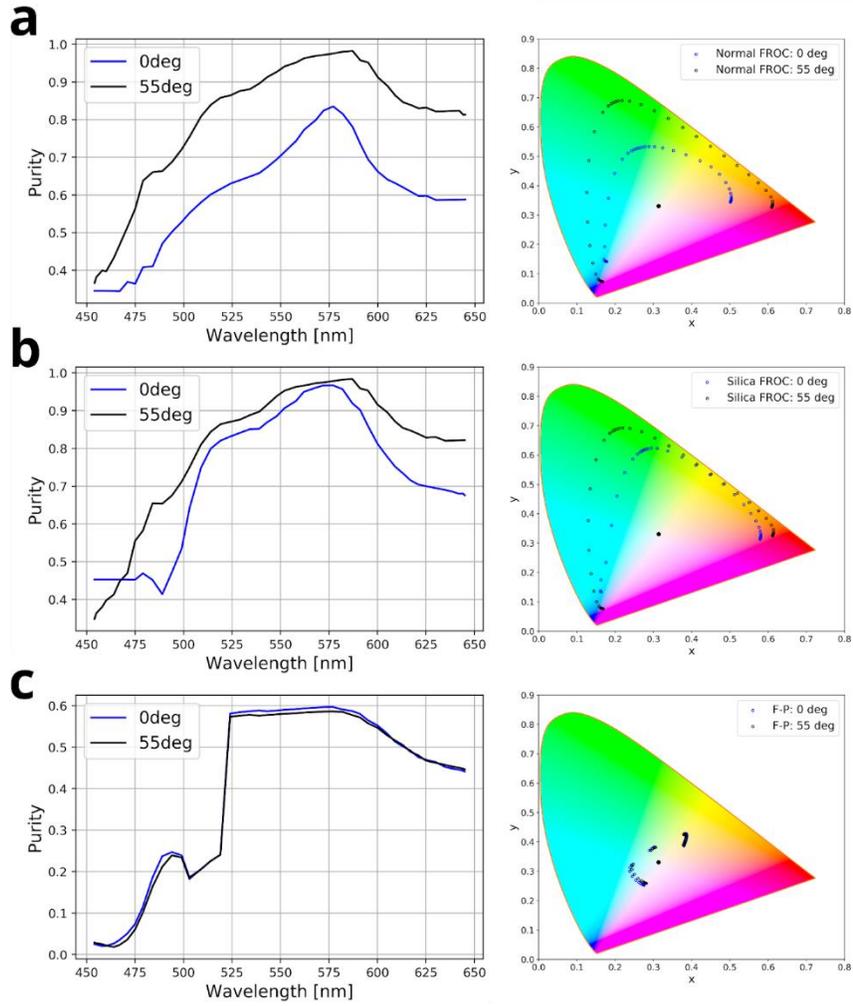

**Figure S5|** Maximum purity for color-optimized FROCs over a range of visible wavelengths. **a** FROCs ( Ge (15 nm)- Ag (x nm)- $TiO_2$(x nm)- Ag (100 nm) ), **b** silica capped FROCs ( with $SiO_2$ (50 nm) ), and **c** Fabry-Perot cavities (Ag (x nm)- $TiO_2$(x nm)- Ag (100 nm) ), for 0° and 55° viewing angles.

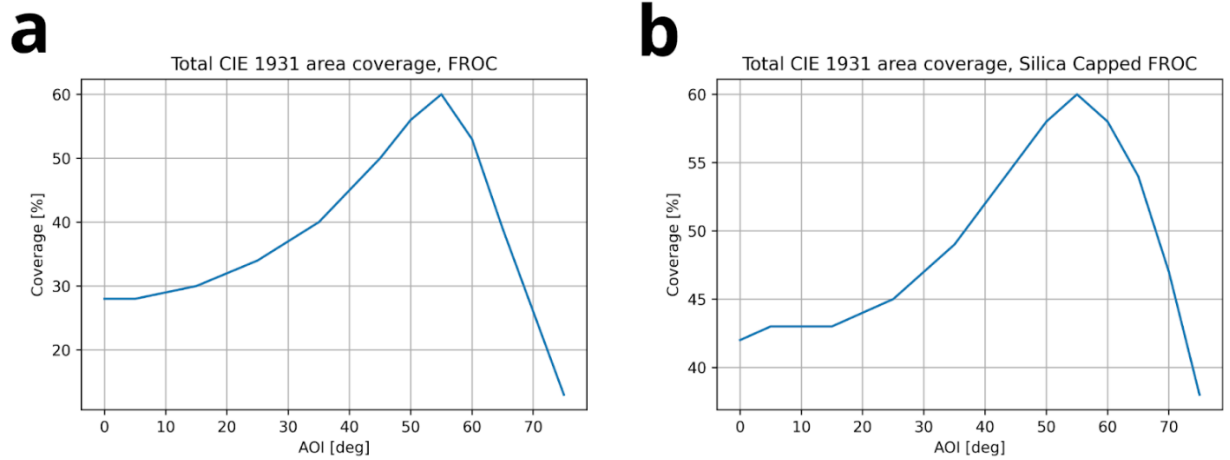

**Figure S6|** CIE 1931 chromaticity space percent coverage for the simulated spectra of **a** FROCs ( Ge (15 nm)- Ag (24 nm)- $TiO_2$ (x nm)- Ag (100 nm) ) and **b** silica capped FROCs ( with $SiO_2$ (50 nm) ) for a range of viewing angles. The thickness of the $TiO_2$ layer is varied. The maximum area coverage occurs at 55° viewing angle. The silica capped FROC has a higher overall CIE area coverage.

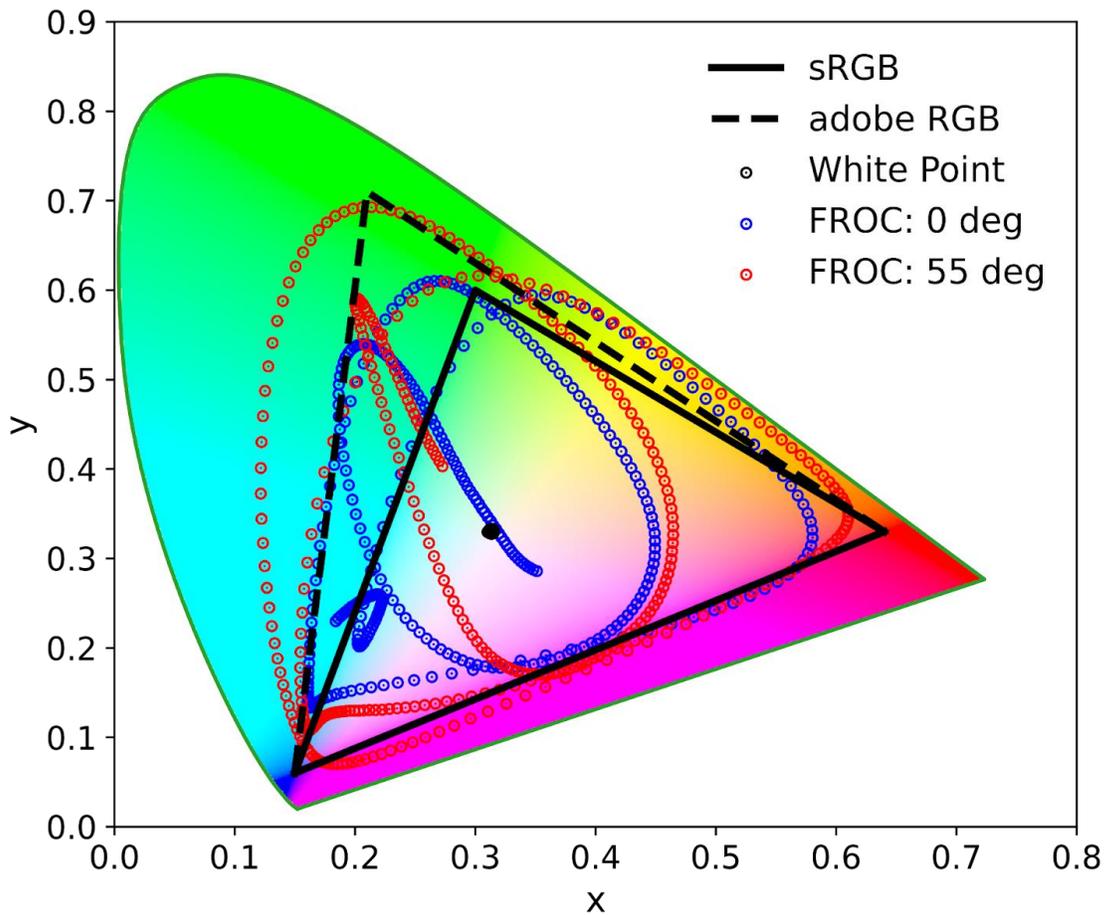

**Figure S7|** CIE 1931 chromaticity coordinates for the simulated spectra of silica capped FROCs with SiO$_2$ (50 nm) -Ge (15 nm)- Ag (24 nm)- TiO$_2$(x nm)- Ag (100 nm) at 0° and 55° viewing angle. The thickness of the TiO$_2$ layer is varied. The sRGB and Adobe RGB color spaces are shown for comparison. FROCs exhibit high coverage of both color spaces through a wide range of viewing angles.